\documentclass[aps,preprint,showpacs,floats,epsf,epsfig,nofootinbib,12pt]{revtex4}
\usepackage{graphicx}
\usepackage{epsfig}
\usepackage[dvips]{color}
\usepackage{subfigure}

\def\beq{\begin{eqnarray}}
\def\eeq{\end{eqnarray}}

\begin{document}

\title{The possible $B\pi$ molecular state  and its radiative decay}

\vspace{1cm}

\author{ Hong-Wei Ke$^1$\footnote{khw020056@hotmail.com}, Lei Gao$^1$ and
        Xue-Qian Li$^2$\footnote{lixq@nankai.edu.cn}  }

\affiliation{  $^{1}$ School of Science, Tianjin University, Tianjin 300072, China \\
  $^{2}$ School of Physics, Nankai University, Tianjin 300071, China }

\vspace{12cm}

\begin{abstract}
Recently, several exotic bosons have been confirmed as multi-quark states, but there
are violent disputes about their inner structures, namely if they are molecular states or
tetraquarks, or even mixtures of the two structures.
It would be interesting to experimentally search for non-strange
four-quark states with open charm or bottom which are lighter than
$\Lambda_c$ or $\Lambda_b$. Reasonable arguments indicate that
they are good candidates of pure molecular states $D\pi$ or $B\pi$
because pions are the lightest boson. Both $B\pi$ and $D\pi$ bound
states  do not decay via strong interaction. The $B\pi$ molecule
may decay into $B^*$ by radiating a photon, whereas $D\pi$
molecule can only decay via weak interaction. In this paper we
explore the mass spectra of $B\pi$ molecular states
 by solving the corresponding B-S
equation. Then the rate of radiative decay
$|\frac{3}{2},\frac{1}{2}\rangle\to B^*\gamma$ is calculated and
our numerical results indicate that the processes can be measured
by the future experiment. We also briefly discuss the $D\pi$ case,
due to the constraint of the final state phase space, it can only
decay via weak interaction.


\end{abstract}

\maketitle

\section{Introduction}
Many charmonium-like or bottomonium-like
resonances $X$, $Y$ and $Z$ bosons, such as
$X(3872)$\cite{Choi:2003ue}, $X(3940)$\cite{Abe:2007jn},
$Y(3940)$\cite{Choi:2005}, $Z(4430)$\cite{Choi:2007wga}
Y(4260)\cite{Aubert:2005rm,Ablikim:2013emm},$Z_c$(4020)\cite{Ablikim:2013wzq},
$Z_c$(3900)\cite{Ablikim:2013mio,Liu:2013dau}, $Z_b(10610)$ and
$Z_b(10650)$ \cite{Collaboration:2011gj} have been experimentally observed.
The data show that there is no room in the
regular representations of $O(3)\otimes SU_f(3)\otimes SU_s(2)$ to
accommodate those newly observed resonances, especially some of them
are charged, thus it is suggested that those exotic bosons are in
multi-quark states.
Since their masses are close
to that of charmonia or bottomonia, thus those states should have hidden
charm or bottom components.  Whereas a newly observed exotic state
$X(5568)$\cite{D0:2016mwd} measured at the $B_s^0 \pi^\pm$
invariant mass spectrum is believed to possess four differently flavored
quarks (antiquarks). If the resonance is eventually confirmed it must be a
four-quark state with open bottom.

Even though so many exotic resonances are confirmed as multi-quark bosons,
there is an acute dispute about their inner structure. By contrast to the
regular quark-anti-quark structure, the system containing two quarks and two anti-quarks
may have different combination patterns: it may reside in a
molecular state, a tetra-quark or a mixing of the two
structures\cite{Xiao:2013iha,Deng:2014gqa,Wang:2014gwa,Wang:2013cya,
Voloshin:2013dpa,Wilbring:2013cha,Cui:2013yva,Zhang:2013aoa,Liu:2013vfa,Dias:2013fza,Ke:2013gia,Burns:2016gvy}.
An intuitive opinion suggests that a narrow-width (i.e. several tens of MeV)
exotic particle might be a molecular state, whereas a wide-width (i.e.
several hundreds MeV) one should be a tetraquark.
However definitely, this naive consideration cannot be a criterion for judging the exotic structure
by merely its width, at most it provides a hint to help confirming the inner structure.
As a matter of fact, so far no any exotic state has ever been firmly determined as a
molecular state.  Actually, if an exotic boson is confirmed to
be in a molecular state, a careful study on it would be very helpful for understanding the dynamics which results in the
different inner structures. Because we lack available data at present, let us theoretically construct
such states which should be ideal  molecular systems.
We would argue that bound states of
$B\pi$ and $D\pi$ should be ideal molecular systems.

The authors of Ref.\cite{Burns:2016gvy} argued that the newly
observed $X(5568)$ contains constituents of $su\bar b\bar d$ which
has an additional valence quark than $\Xi_b$ with $usb$ contents,
has mass of 5619.5 MeV, it is lighter than the mass of $\Xi_b$, so
that $X(5568)$ seems not to be a tetraquark of $su\bar b\bar d$,
if it indeed exists. The other research
works\cite{Guo:2016nhb,Lu:2016zhe,Chen:2016npt} support that such
tetraquark with constituents of $su\bar b\bar d$ should be heavier
than $\Xi_b$.

Following this argument we would be tempted to suppose that if a
non-strange four-quark state
with open bottom or charm  exists and is
lighter than $\Lambda_b$ or $\Lambda_c$, the only possible choice
is that they are pure molecular states of  $B\pi$ or $D\pi$. The
reason is that pion is the lightest boson and especially lighter
than a valence quark. Even though the reason why pions are so
light is still a not fully understood enigma yet, the fact that it
is lighter than valence quarks is surely confirmed. More
concretely, since the mass of $\pi$ is lighter than any
constituent quark generally the molecular state of $B\pi$ or
$D\pi$ should be lighter than the tetraquark state with the same
quark-structure and as well as the corresponding baryons such as
$\Lambda_b$ or $\Lambda_c$. Namely, we are going to experimentally
search for exotic four-quark states which are lighter than
$\Lambda_b$ or $\Lambda_c$ because there is a strong evidence that
they are hadronic molecules. Moreover, if the bound state of
$B\pi$ ($D\pi$) is experimentally confirmed, we will have all
reasons to believe that other molecular states indeed exist in
nature, that is why exploration of $B\pi$  and $D\pi$ bound states
is crucially important.

Obviously, molecules $B\pi$ or $D\pi$ do not decay via
strong interactions, therefore, one expects to observe them only at
radiative and/or weak processes. It would definitely make
detection more difficult, but not impossible.
Indeed the bound state $B\pi$  may decay into $B^*\gamma$ with a larger
rate than  weak  decay modes. Thus
we will more focus on the $B\pi$ bound state and its
radiative decay in this paper.

In the quantum field theory at the lowest order two particles
interact with each others by exchanging certain particles.  For
our case, the molecular state consists of two  color-singlet
mesons, we can derive the effective hamiltonian which corresponds
to exchanging scalar (such as $\sigma$) or vector (such as $\rho$)
etc.) mesons between $B$ and $\pi$ (or $D$ and $\pi$).

In Ref.\cite{Guo:2007mm} the authors employed the Bethe-Salpeter
equation to study the $K\bar K$ or $BK$ molecular state and their
decays. In this work we follow their approach to study the
molecular state of $B\pi$. Here we only concern the ground
states i.e.  the orbital angular momentum between the two constituent
mesons is zero ($l=0)$ so the $J^{P}$ of the molecular state is
$0^+$. Since the isospins of  $B$ and $\pi$ are $1/2$ and 1, the
isospin state of $B\pi$  can reside in either 3/2 or 1/2 states.
Different isospin states have different effective vertices for the strong-interaction
which determine if the bound sates can be formed as a physical object. We will solve
the B-S equation first to explore the possibility of forming the
bound state and obtain the corresponding B-S wave function. Then with the wave
function we estimate its radiative decay rate in the same framework.

After this introduction we will present the B-S equations for the
$0^+$ molecular state and derive the formula for its radiative
decay rate. Then in section III we present our numerical results
along while explicitly displaying all input parameters. Section IV
is devoted to a brief summary. As we indicated above, in this work
we concentrate on the case of $B\pi$ molecular states, then in the
last section, we will briefly discuss the $D\pi$ case.

\section{The bound states of $B\pi$ and their radiative decay in the Bethe-Salpeter framework }
\begin{figure*}
        \centering
        \subfigure[~]{
          \includegraphics[width=7cm]{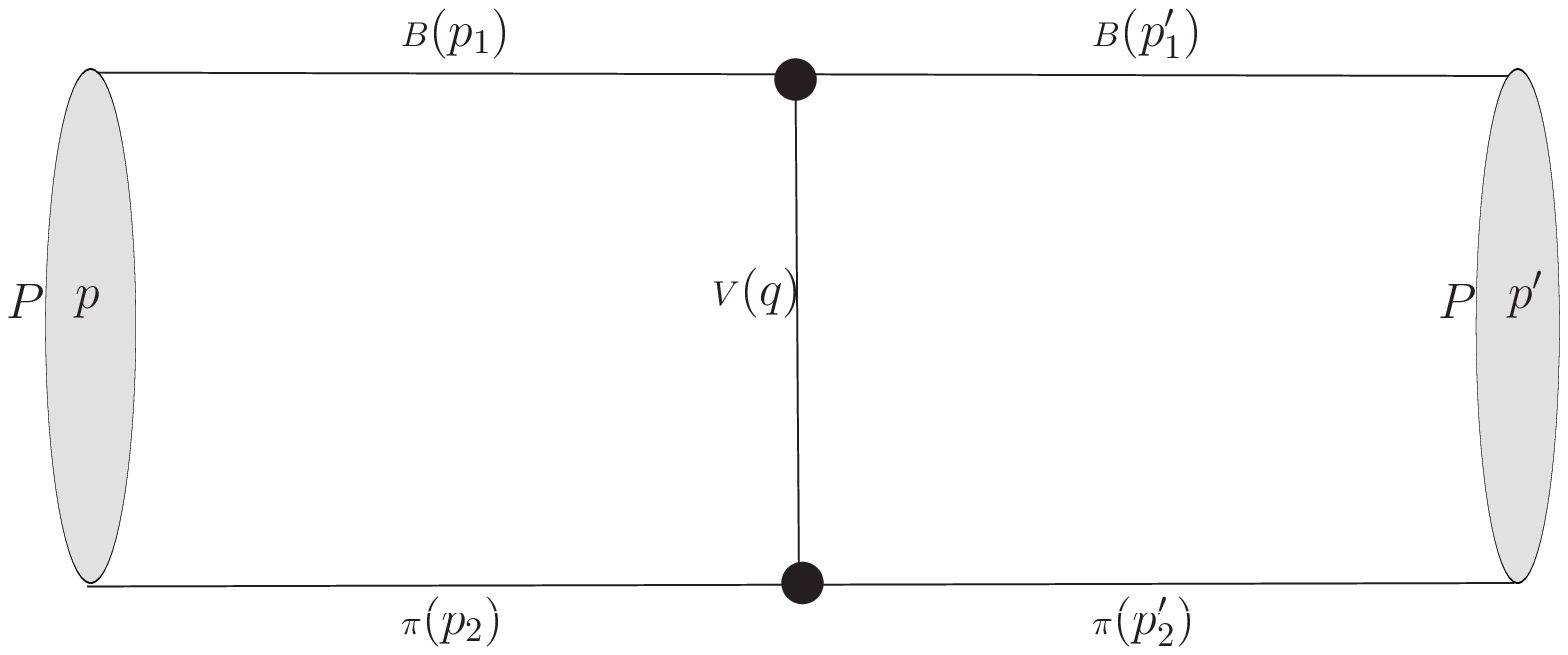}}
        \subfigure[~]{
          \includegraphics[width=7cm]{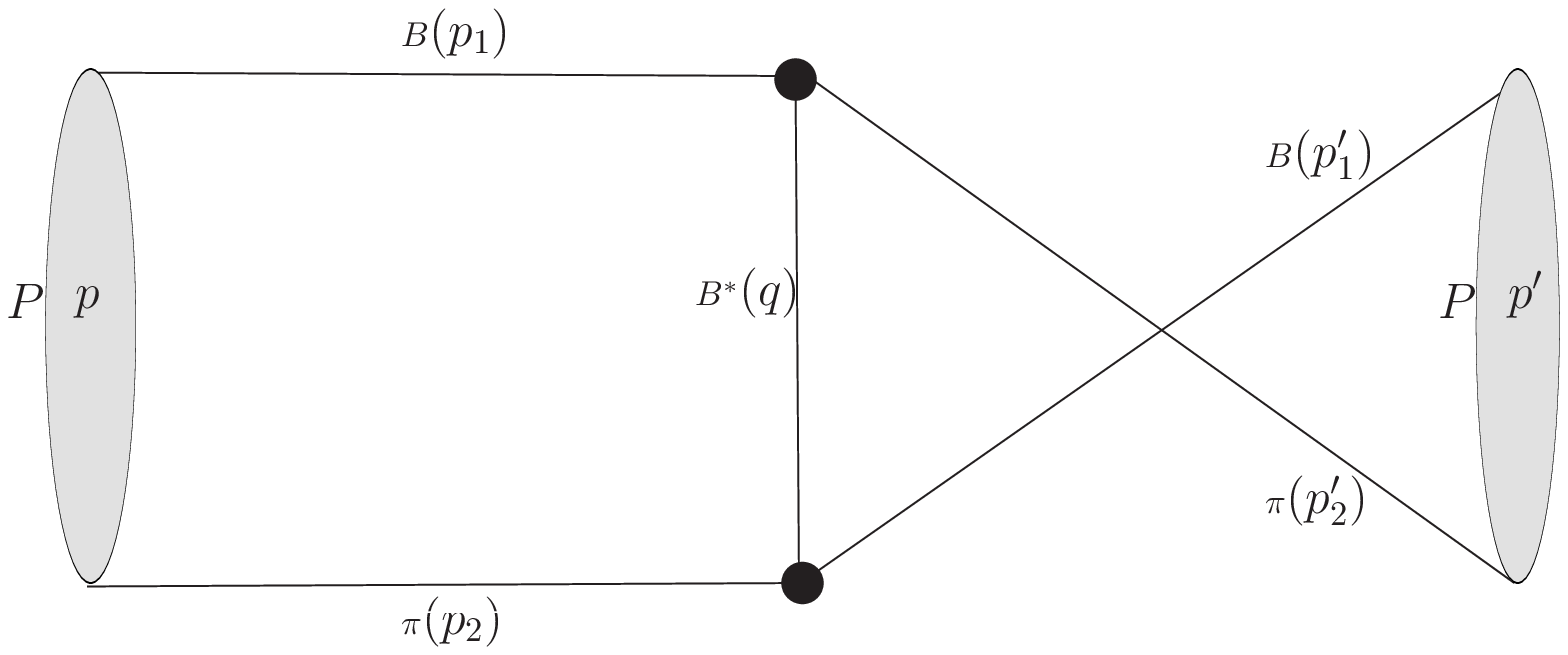}}
 \caption{the bound states of  $B\pi$ formed by exchanging light vector mesons (a) and $B^*$ (b) .}
        \label{decay1}
    \end{figure*}

\subsection{the molecular state of $B\pi$}
Since the isospins of $B$ and $\pi$ are 1/2 and 1  the possible
bound states of $B$ $\pi$
should be in two isospin assignments i.e. $|I,I_3\rangle$ are $|\frac{1}{2},\pm\frac{1}{2}\rangle$
 $|\frac{3}{2},\pm\frac{1}{2}\rangle$ and $|\frac{3}{2},\pm\frac{3}{2}\rangle$. Let us work on the isospin states
$$|\frac{1}{2},\frac{1}{2}\rangle=\sqrt{\frac{2}{3}}|B^0\pi^+\rangle-\sqrt{\frac{1}{3}}|B^+\pi^0\rangle,$$
,
$$|\frac{3}{2},\frac{1}{2}\rangle=\sqrt{\frac{1}{3}}|B^0\pi^+\rangle+\sqrt{\frac{2}{3}}|B^+\pi^0\rangle.$$
and
$$|\frac{3}{2},\frac{3}{2}\rangle=|B^+\pi^+\rangle.$$
While the states $|\frac{1}{2},-\frac{1}{2}\rangle$,
$|\frac{3}{2},-\frac{1}{2}\rangle$  and
$|\frac{3}{2},-\frac{3}{2}\rangle$ are just the charge conjugate
states of  $|\frac{1}{2},\frac{1}{2}\rangle$,
$|\frac{3}{2},\frac{1}{2}\rangle$  and
$|\frac{3}{2},\frac{1}{2}\rangle$, therefore their properties are the same.

\subsection{The Bethe-Salpeter (B-S) equation for $0^+$ molecular state}
Two mesons may form a bound state by exchanging appropriate mesons, and the scenario
is depicted in Fig.\ref{decay1}. The relative  and
total momenta of the bound state in the equations are defined
as
\begin{eqnarray} p = \eta_2p_1 -
\eta_1p_2\,,\quad p' = \eta_2p'_1 - \eta_1p'_2\,,\quad P = p_1 +
p_2 = p'_1 + p'_2 \,, \label{momentum-transform1}
\end{eqnarray}
where $p$ and $p'$ are the relative momenta before and after the effective vertices,
$p_1$ ($p'_1$) and $p_2$ ($p'_2$) are those momenta of the
constituents before  and after the effective vertices, $P$ is the total
momentum of the bound state, $\eta_i = m_i/(m_1+m_2)$ and $m_i\,
(i=1,2)$ is the mass of the $i$-th constituent meson.

The corresponding B-S equation was deduced in Ref.\cite{Guo:2007mm,Feng:2011zzb} as
\begin{eqnarray} \label{3-dim-BS1}
{E^2-(E_1+E_2)^2\over (E_1+E_2)/E_1E_2}
\widetilde\chi_{{}_\mathcal{P}}^{}({\bf p}) ={i\over
2}\int{d^3\mathbf{p}'\over(2\pi)^3}\, {\overline{} }K({\bf p},{\bf
p}')\widetilde\chi_{{}_\mathcal{P}}^{}({\bf p}')F(\bf p-\bf p')^2
\,,
\end{eqnarray}
where $E$ is the total energy of the bound state, $E_i =
\sqrt{{\bf p}^2 + m_i^2}$ and
$\widetilde\chi_{{}_\mathcal{P}}^{}({\bf p})$ is the B-S wave
function in the three-momentum space. Therefore, the key point is
to determine the kernel function $K({\bf p},{\bf p}')$.

Since the constituent
mesons are not point particles,  a form factor at each effective
vertex should be introduced to reflect the finite-size
effects of these hadrons. The form factor is assumed to be in the form:
\begin{eqnarray} \label{form-factor} F({\bf k}) = {2\Lambda^2 -
M_{\rm V}^2 \over 2\Lambda^2 + {\bf k}^2}\,,\quad {\bf k} = {\bf
p}-{\bf p}' \,,
\end{eqnarray}
where $\Lambda$ is a cutoff parameter and usually fixed by fitting data.
For  exchanging a light vector ($\rho$ or $\omega$) between the mesons  as shown in
Fig.\ref{decay1}(a), the kernel is
\begin{eqnarray} K({\bf p},{\bf
p}') = i C_{I,I_3}\,{g_{_{BBV}}g_{_{ V\pi\pi}}'}\, {({\bf p}+{\bf
p}')^2 + 4\eta_1\eta_2 E^2 + ({\bf p}^2-{\bf p}'{}^2)^2/M_{\rm
V}^2 \over ({\bf p}-{\bf p}')^2 + M_{\rm V}^2} \,.
\label{potential-with-isospin-factor1}
\end{eqnarray}
The Feynman diagram for exchanging $\sigma(f_0(500))$ is the same
as in Fig.\ref{decay1}(a), the kernel is
\begin{eqnarray} K({\bf p},{\bf
p}') = i
C_{I,I_3}\,4m_Bm_\sigma{g_{_{\sigma}}g_{_{\sigma\pi\pi}}}\, {1
\over ({\bf p}-{\bf p}')^2 + M_{\rm \sigma}^2} \,.
\label{potential-with-isospin-factor2}
\end{eqnarray}
While for exchanging $B^*$  the kernel (shown in
Fig.\ref{decay1}(b)) is
\begin{eqnarray} K({\bf p},{\bf
p}') = -i C_{I,I_3}\,{g_{_{B^*BB}}g_{_{B^*\pi\pi}}}\, {({\bf
p}-{\bf p}')^2 + E^2 + [\eta_1E^2-({\bf p}^2-{\bf p}'{}^2)]
[\eta_1E^2+({\bf p}^2-{\bf p}'{}^2)]/M_{ B^*}^2 \over ({\bf
p}+{\bf p}')^2 + M_{ B^*}^2-(\eta_1-\eta_2)^2E^2} \,.
\label{potential-with-isospin-factor3}
\end{eqnarray}

Since the function
$\widetilde\chi(\mathbf{p})$ only depends on the norm of the
three-momentum we may first integrate over the azimuthal angle
in Eq. (\ref{3-dim-BS1})
$$\frac{i}{2}\int{d^3\mathbf{p}'\over(2\pi)^3}\, {\overline{} K}({\bf p},{\bf
p}')F(\bf p-\bf p')^2,  $$ to obtain a new form
$U(|\mathbf{p}|,|\mathbf{p}'|)$
corresponding to Eq.
(\ref{potential-with-isospin-factor1}), and it can be found in
Ref.\cite{Guo:2007mm}. Then the B-S equation turns into a simplified
one-dimension integral equation
\begin{eqnarray} \label{3-dim-BS4}
\widetilde\chi({\bf |p|}) ={(E_1+E_2)/E_1E_2\over E^2-(E_1+E_2)^2
}\int{d \mathbf{|p}'|}\, {\overline{} U}({\bf |p|},{\bf
|p}'|)\widetilde\chi({\bf |p}'|) .
\end{eqnarray}

In terms of the approach given in
Ref.\cite{Guo:2007mm,Feng:2011zzb}  the isospin factor can be
obtained. For the $B^0\pi^+$ molecule, the corresponding isospin
factor $C_{I.I_3}$ appearing in
Eqs.(\ref{potential-with-isospin-factor1}) and (\ref
{potential-with-isospin-factor2}) takes different values as
$C_{\frac{1}{2},\frac{1}{2}}$ are $1-\sqrt{2}$ , $1$, $1$ and
$2-\sqrt{2}$  whereas $C_{\frac{3}{2},\frac{1}{2}}$ are
$1+2\sqrt{2}$, $1$, $1$ and $2+2\sqrt{2}$ corresponding to
respectively exchanging $\rho$, $\omega$, $\sigma$ and $B^*$.
Whereas for the $B^+\pi^0$ molecule, the isospin factor changes as
$C_{\frac{1}{2},\frac{1}{2}}$ being $1-2\sqrt{2}$, $1$, $1$ and
$-2\sqrt{2}$; $C_{\frac{3}{2},\frac{1}{2}}$ are $1+\sqrt{2}$, $1$,
1 and $2+\sqrt{2}$, instead. Since the values of $C_I$ are
different for $B^0\pi^+$ and $B^+\pi^0$ we will solve their B-S
equations respectively. For the $B^+\pi^+$ system the isospin
factor $C_{\frac{3}{2},\frac{3}{2}}$ would take 1,1,2 and 1
corresponding to exchanging three different vector mesons $\rho$,
$\omega$ and $B^*$ and $\sigma$.

In order to employ the wave function one first needs to normalize it. The
normalization condition is
\begin{eqnarray} -\frac{1}{\pi^2}\int\frac{d^3\mathbf{p}}{(2\pi)^3}
\chi_{{}_\mathcal{P}}^{}({\bf p})^2R-\frac{1}{4\pi^2}\int
\frac{d^3\mathbf{p}d^3\mathbf{p'}}{(2\pi)^6}\chi_{{}_\mathcal{P}}^{}({\bf
p})\chi_{{}_\mathcal{P}}^{}({\bf
p}')F(\mathbf{p}-\mathbf{p'})\frac{\partial
K(\mathbf{p},\mathbf{p'})}{\partial E}=1.
\end{eqnarray}
where
\begin{eqnarray}
R=&&-E[-2E^2(E_1^2-E_2^2)(E_1\eta_1-E_2\eta_2)+E^4(E_1\eta_1+E_2\eta_2)\nonumber\\&&
+(E_1^2-E_2^2)(E_1^3\eta_1+3E_1E_2^2\eta_1-3E_1^2E_2\eta_2-E_2^3\eta_2)]
\nonumber\\&&\frac{1}{{2E_1E_2[E^4+(E_1^2-E_2^2)^2-2E^2(E_1^2+E_2^2)]^2}}.
\end{eqnarray}


\subsection{Estimating the decay rate of $B\pi$ molecule to
$B^*+\gamma$}
\begin{figure*}
        \centering
        \subfigure[~]{
          \includegraphics[width=7cm]{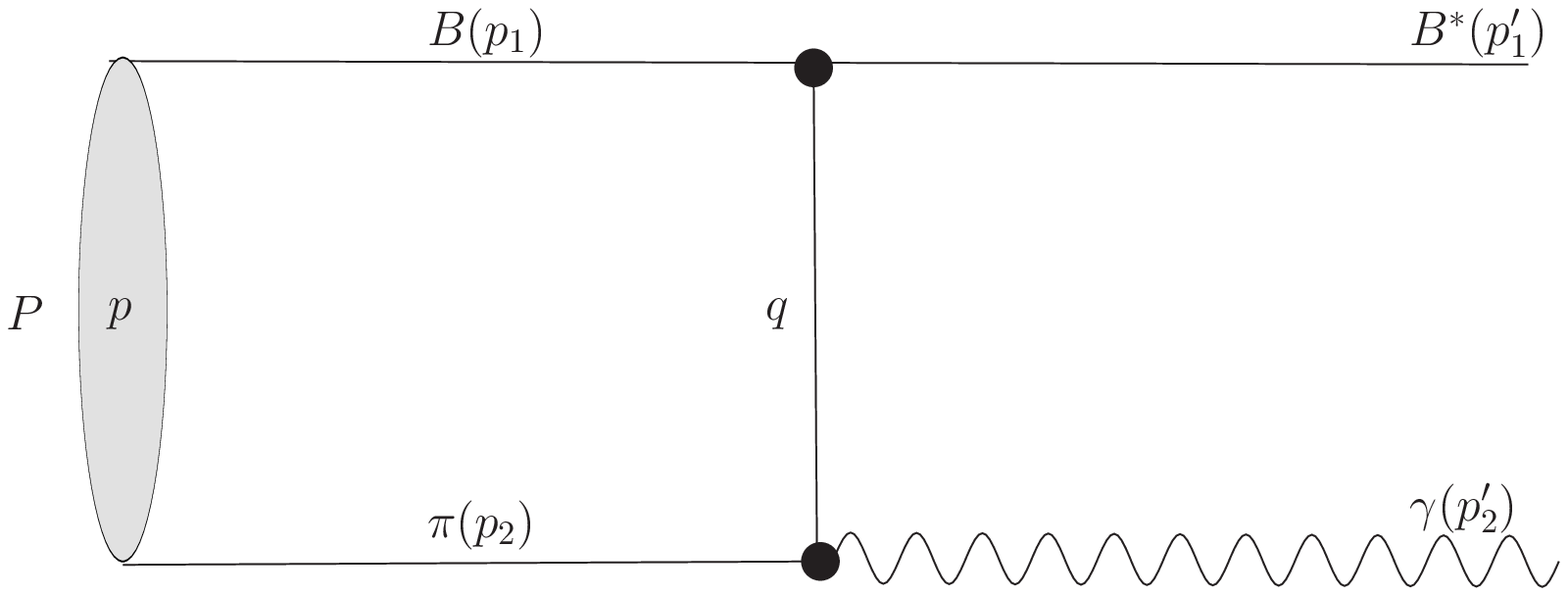}}
        \subfigure[~]{
          \includegraphics[width=7cm]{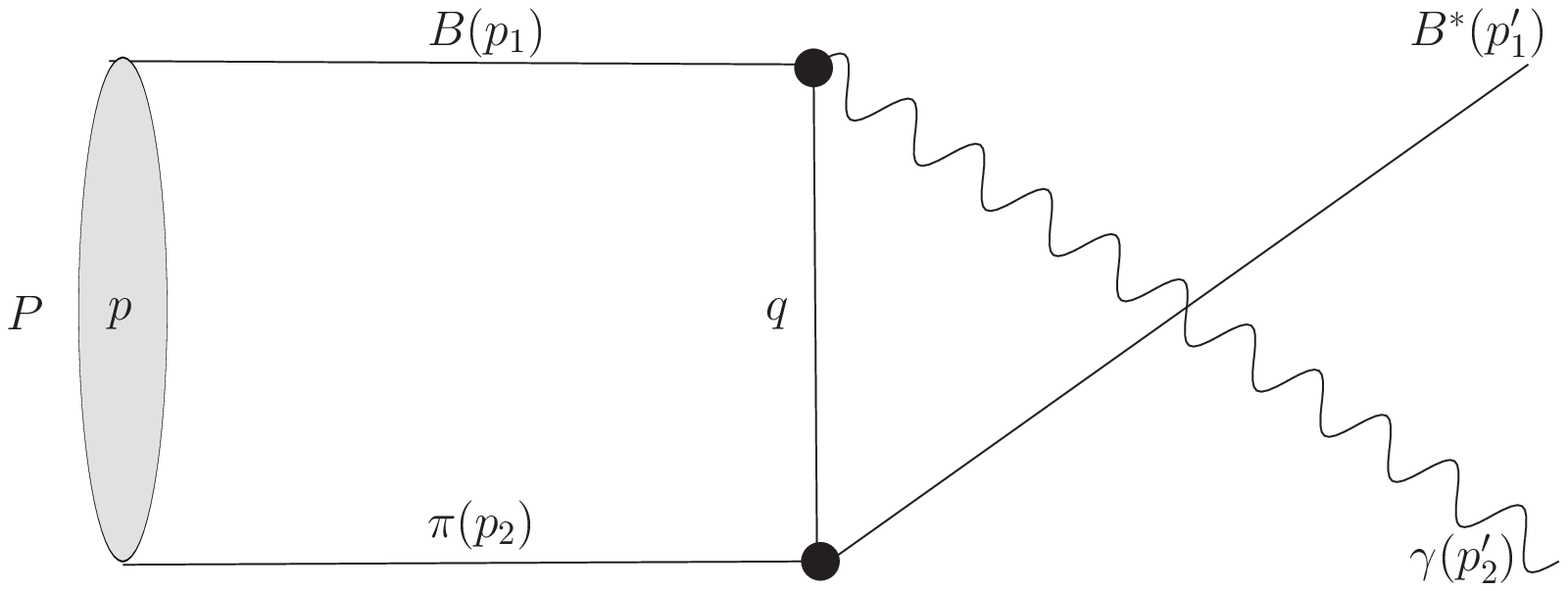}}
 \caption{The radiative decay of the bound state}
        \label{decay}
    \end{figure*}

It is crucial to ask a question, that as $B$ and $\pi$ constitute
a bound state, i.e. a hadronic molecule, how
can we identify the molecular four-quark system? As usual, to
confirm the inner structure, one needs to measure its spectrum via its production
and decay patterns.
$B$ and $\pi$ constitute a ground state hadronic molecule, which
cannot decay via strong interaction. Actually,
the overwhelming decay portals of these bound states are induced by weak interaction, whose rates
are small, so experimental detections are rather
difficult, because the complex background, especially for the hadron colliders.
Fortunately some molecular
states can decay by radiating a photon, obviously such process is easier to be
observed and our $B\pi$ molecule is just the case.

The Feynman
diagrams for radiative decays of the $B\pi$ molecule are shown in Fig.\ref{decay}.
Fig.\ref{decay} (a) corresponds to exchanging $\rho$, $\omega$ or $\pi$ while
Fig.\ref{decay} (b) is for exchanging $B$ or $B^*$. Following
Ref.\cite{Guo:2007mm,Feng:2011zzb,Ke:2012gm} the
transition matrix elements by exchanging $\rho(\omega)$, $\pi$,
$B$ and $B^*$ are
\begin{eqnarray}\label{m1}
M_{\rho(\omega)}&&=i\frac{\sqrt{2E}g_{BB^*\rho}g_{\rho\pi\gamma}}{m_{B^*}}C'_{I,I_3}\int\frac{d^4p}{(2\pi)^4}q_c{p_1'}_a{\epsilon_1}_b\varepsilon^{abc\mu}
({p_2}-q)_\sigma{p_2'}_\alpha{\epsilon_2}_\beta
\varepsilon^{\alpha\beta\nu\sigma}\frac{g_{\mu\nu}-q_\mu
q_\nu/M_\rho^2}{M^2_\rho-q^2}F(|\mathbf{q}|)^2\chi(p)\\M_{\pi}&&=i\sqrt{2E}g_{BB^*\pi}g_{\pi\pi\gamma}C'_{I,I_3}\int\frac{d^4p}{(2\pi)^4}
4q_bq_\beta{\epsilon_1}_b{\epsilon_2}_\beta\frac{g_{\mu\nu}-q_\mu
q_\nu/M_\pi^2}{M^2_\pi-q^2}F(|\mathbf{q}|)^2\chi(p)
\\M_{B^*}&&=-i\frac{\sqrt{2E}g_{BB^*\gamma}g_{B^*B^*\pi}}{m_{B^*}}C'_{I,I_3}\int\frac{d^4p}{(2\pi)^4}
{p_2'}_a(p_1+q)_c{\epsilon_2}_\beta\varepsilon^{a\beta \mu
c}{p_2}_\sigma({p_1'}+q)_\alpha{\epsilon_1}_b
\\&&\varepsilon^{\alpha b\nu\sigma}\frac{g_{\mu\nu}-q_\mu
q_\nu/M_B^{*2}}{M^2_{B^*}-q^2}F(|\mathbf{q}|)^2\chi(p)
\nonumber\\M_{B}&&=i\sqrt{2E}g_{BB\gamma}g_{B^*B\pi}C'_{I,I_3}\int\frac{d^4p}{(2\pi)^4}
4q_bq_\beta{\epsilon_1}_b{\epsilon_2}_\beta\frac{g_{\mu\nu}-q_\mu
q_\nu/M_B^2}{M^2_B-q^2}F(|\mathbf{q}|)^2\chi(p)
\end{eqnarray}
where $\epsilon_1$ and $\epsilon_2$ are the polarizations of $B^*$
and photon respectively. For $B^0\pi^+$ the isospin factor
$C_{\frac{3}{2},\frac{1}{2}}$ takes
$\frac{2}{\sqrt{3}},\;0,\;\frac{2}{\sqrt{3}},\;\frac{1}{\sqrt{3}},\;\frac{1}{\sqrt{3}}$
corresponding to exchanging $\rho$, $\omega$, $\pi$, $B$ and
$B^*$, whereas for $B^+\pi^0$ the isospin factor
$C_{\frac{3}{2},\frac{1}{2}}$ are respectively
$\sqrt{\frac{2}{3}},\sqrt{\frac{2}{3}},\;
\sqrt{\frac{2}{3}},\;2\sqrt{\frac{2}{3}},\; 2\sqrt{\frac{2}{3}}$.
To simplify our calculation we set $p_0=0$ in the kinetic part
of the integrand in Eqs.(9), for example, in
Eq.(10), $p_0=0$ applies merely to
$q_c{p_1'}_a{\epsilon_1}_b\varepsilon^{abc\mu}({p_2}-q)_\sigma{p_2'}_\alpha{\epsilon_2}_\beta
\varepsilon^{\alpha\beta\nu\sigma}\frac{g_{\mu\nu}-q_\mu
q_\nu/M_V^2}{M_V-q^2}$, then the integrand turns into
\begin{eqnarray}\label{m2}
M_\rho=i\sqrt{2E}g_1g_2C'_{I,I_3}\int\frac{d^3p}{(2\pi)^3}q_c{p_1'}_a{\epsilon_1}_b\varepsilon^{abc\mu}({p_2}-q)_\sigma{p_2'}_\alpha{\epsilon_2}_\beta
\varepsilon^{\alpha\beta\nu\sigma}\frac{g_{\mu\nu}-q_\mu
q_\nu/M_\rho^2}{M^2_\rho-q^2}F(|\mathbf{q}|)^2\tilde{\chi}(\mathbf{p}).
\end{eqnarray}
where the definition $ \widetilde\chi_{_P}({\bf p})= \int dp^0 \,
\chi_{_P}(p) \, $ is used. Namely, in the new expression, the
argument of $\widetilde\chi_{_P}({\bf p})$ is a three-momentum
${\bf p}$ instead of the four momentum $p$. It is noted that this
simplification is similar to the instantaneous approximation for
solving the B-S equation which is usually adopted.

Generally we can define two form
factors for the transition
 \begin{eqnarray}
 M=F_1\epsilon_1\cdot\epsilon_2+F_2\epsilon_1\cdot P\epsilon_2\cdot
 P
\end{eqnarray}
and $F_1$ and $F_2$ can be extracted from Eq .(\ref{m2}) and
calculated numerically.

\section{numerical results}
To solve the B-S equation and numerically calculate the radiative
decay rate some input parameters are needed. The mass of $B$,
$B^*$, $\rho$, $\omega$, $\pi$ are taken from the
databook\cite{PDG12}.

On the other hand we need to determine the relevant coupling
constants appearing at the effective vertices. By calculating the
transition $\rho\to\pi\pi$ and comparing the result with the
data\cite{PDG12} one can fix the coupling $g_{\rho\pi\pi}=5.97$.
Similarly we fix $g_{\omega\pi\pi}=0.175$,
$g_{\rho\pi\gamma}=0.417$GeV$^{-1}$,
$g_{\omega\pi\gamma}=1.215$GeV$^{-1}$. However for determining the coupling
constants involving $B^{(*)}$ mesons there are not available data,
so we fix $g_{D^*D\pi}=8.05$ and $g_{D^*D\gamma}=0.706$GeV$^{-1}$
by using relations $g_{B^*B^*\pi}=g_{B^*B\pi}=g_{D^*D\pi}$ and
$g_{B^*B^*\gamma}=g_{B^*B\gamma}=g_{D^*D\gamma}$ which are
reasonable in the heavy quark limit. $g_{BB\rho}=g_{B^*B\rho}=3$
is taken from Ref.\cite{Ke:2012gm}. The $g_{\rho}=0.76$ was fixed
in Ref.\cite{Lee:2009hy}. If one sets $m_\sigma=500$ MeV and $\Gamma_\sigma=550$ MeV,
$g_{\sigma\pi\pi}=4.09$ is obtained. $\Lambda$ is the cutoff parameter which will
be used while searching for a solution of the B-S equation.
In Ref.\cite{Meng:2007cx} the value of $\Lambda$ is suggested to be
0.88 GeV to 1.1 Gev. In this work
letting  $\Lambda$ span in the range from $0.8$
to $1.2$ GeV, we solve the B-S equation.

\begin{figure}
\begin{center}
\begin{tabular}{ccc}
\scalebox{0.8}{\includegraphics{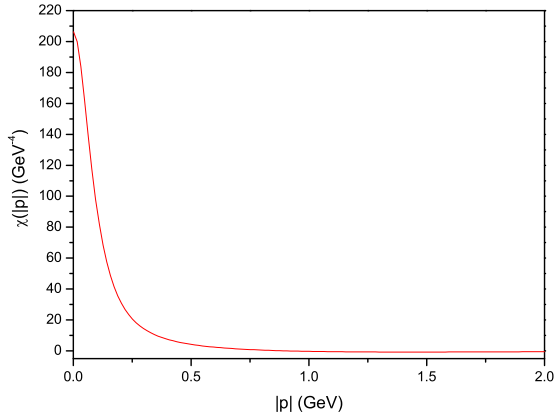}}
\end{tabular}
\end{center}
\caption{the B-S wave function of the molecular state of $B^0\pi^+$
($\Delta E=40$ MeV)}\label{wave}
\end{figure}


\begin{table}[!h]
\caption{The  $\Lambda$ for different bound energies of
$B\pi(I=\frac{3}{2}, I_{z}=\frac{1}{2}$)  .}\label{Tab:t1}
\begin{ruledtabular}
\begin{tabular}{ccccccccc}
  $\Delta$E(MeV)    & -10
&-20 &  -30&-40    & -50 & -60  &  -70&-80\\\hline $\Lambda
(B^0\pi^+)$&0.991 &1.024& 1.053& 1.080& 1.102& 1.122& 1.140&
1.155\\\hline $\Lambda(B^+\pi^0)$&0.947 &0.979& 1.006& 1.030&
1.051& 1.068& 1.084& 1.096
\end{tabular}
\end{ruledtabular}
\end{table}

\begin{table}[!h]
\caption{The  $\Lambda$ for different bound energies of
$B^+\pi^+(I=\frac{3}{2}, I_{z}=\frac{3}{2}$) .}\label{Tab:t2}
\begin{ruledtabular}
\begin{tabular}{ccccccccc}
  $\Delta$E(MeV)    & -10
&-20 &  -30&-40    & -50 & -60  &  -70&-80\\\hline
$\Lambda$&0.878& 0.906& 0.930& 0.950& 0.967& 0.981& 0.993& 1.003
\end{tabular}
\end{ruledtabular}
\end{table}
We now solve the B-S equation. $\bf |p|(|p'|)$  takes
$n$ discrete values which are arranged in order from small to large and the gap between two adjacent
values is $\Delta \bf p$, then $\chi({\bf |p|})$ can constitute
a column matrix and the coefficients on the right side of Eq.
(\ref{3-dim-BS4}) make an $n\times n$ matrix $M$.
Our strategy is following. The binding energy is  $\Delta E=m_1+m_2-E$, thus we write up the determinant of
$M(\Delta E,\Lambda)-I$ ($I$ is a unit matrix) where $M(\Delta E,\Lambda)$ is a matrix function of
the binding energy $\Delta E$
and parameter $\Lambda$. Then setting equation $|M(\Delta E,\Lambda)-I|=0$ which is equivalent to
the secular equation in regular quantum mechanics, by varying $\Delta E$ we obtain a series of solutions
for $\Lambda$. We would check whether the obtained values of $\Lambda$ fall within the range of 0.8 to 1.2 GeV
which is priori set. If the answer is yes, we would conclude that the bound state should exist.  Moreover, with
the obtained $\Delta E$ and $\Lambda$, the B-S wave function is achieved.

When we try to solve the B-S equation for the $B\pi$ system in isospin
$|\frac{1}{2},\frac{1}{2}\rangle$ state, we find that
by setting different binding
energies one cannot achieve a value of $\Lambda$ which falls in the supposed range
0.8-1.2 GeV, so that we would determine that a $B\pi$  bound state of isospin $(1/2,1/2)$ does not exist in
Nature. By contrary, the isospin $|\frac{3}{2},\frac{1}{2}\rangle$ $B\pi$
bound state does exist. According to the aforementioned  $C_{I,I_3}$ values
one can understand that the
interaction between $B$ and $\pi$ in
$|\frac{1}{2},\frac{1}{2}\rangle$ system is not strong enough to
bind the constituents but it is sufficiently large for
$|\frac{3}{2},\frac{1}{2}\rangle$. In table \ref{Tab:t1} we
present the $\Lambda$ values for the bound state of
 $B\pi$ state $|\frac{3}{2},\frac{1}{2}\rangle$. The normalized wave
function is depicted in Fig. \ref{wave}. For the bound state
$B^+\pi^+$ besides the strong interaction there electromagnetic interaction
also applies, however comparing the electromagnetic coupling $e^2$ with the effective
strong coupling $g_1g_2$, one can safely ignore the contribution of  electromagnetic interaction
at all.

On other aspect, even though $|\frac{3}{2},\frac{1}{2}\rangle$  bound states do not decay via
strong interaction, they decay into other hadrons by emitting a photon, i.e  a
radiate decay.  The form factor $F_1$ and $F_2$ in the transition
$M\rightarrow B^*\pi$ are calculated numerically. The theoretically estimated decay rates
are present in table \ref{Tab:t3} for different binding energies.

\begin{table}[!h]
\caption{The form factors and decay widths for different binding
energies of $B^0\pi^+(I=\frac{3}{2}, I_{z}=\frac{1}{2}$)
.}\label{Tab:t3}
\begin{ruledtabular}
\begin{tabular}{ccccccccc}
  $\Delta$E(MeV)    & -10
&-20 &  -30&-40    & -50 & -60  &  -70&-80\\\hline
$F_1$(GeV)&-0.287 & -0.472  & -0.630  &-0.790 &-0.971 &-1.176
&-1.455&-1.800
\\\hline $F_2$(GeV)$^{-1}$&0.0127 &0.0238  & 0.0364&0.0525   &
0.0738 &-0.0966 &-0.140  &-0.194\\\hline $\Gamma$(keV)&2.79 & 7.07
&12.48&19.98&30.65
& 40.09 &59.97 &68.12\\
\end{tabular}
\end{ruledtabular}
\end{table}

\section{conclusion and discussions}

In this work we study the bound state of $B\pi$ which seems to
be identified as a pure molecular state and meanwhile as long as it
is experimentally observed, we can firmly determine existence of hadronic
molecules. Combining future experimental data with the results provided in this work, we would gain valuable
information about the structures of the four-quark exotic states and moreover, the
applying dynamics.

We suggest that  by solving
the B-S equation  with appropriate effective interaction between two constituent hadrons
one can determine whether the four-quark system can be bound as a
molecular state. Since the constituents are hadrons
the effective interactions can be derived in terms of field theory. Since the
isospins of $B$ and $\pi$ are 1/2 and 1 respectively the bound
state can be $|\frac{1}{2},\frac{1}{2}\rangle$,
$|\frac{3}{2},\frac{1}{2}\rangle$ and
$|\frac{3}{2},\frac{3}{2}\rangle$. Priori setting a reasonable range for
the parameter $\Lambda$ within 0.8-1.2 GeV according to the suggestions given in literature,
one can numerically solve the B-S equation to gain the binding energies and wavefunctions of the systems
with quantum number $|\frac{3}{2},\frac{1}{2}\rangle$
and $|\frac{3}{2},\frac{3}{2}\rangle$. Our numerical results show that there is
not a solution for the bound state with quantum number
$|\frac{1}{2},\frac{1}{2}\rangle$.

Since the parameter cannot be determined very precisely, our
prediction on the mass spectrum of the bound state is also not
very accurate as the errors come with uncertainties of theoretical inputs. As solving the
B-S equation of the system for different binding energies, the corresponding parameter $\Lambda$ and B-S
wave function are obtained. With the wave function we can
estimate the radiative decay rate of
$B\pi(|\frac{3}{2},\frac{1}{2}\rangle)\rightarrow B^*\gamma$.
It is found that the partial width can vary in a
certain range with different input values of $\Lambda$. We lay hope on the future measurement
which will tell us the measured values of the binding
energy and partial width of radiative decay. The data  would check our calculation
and help to fix the relevant parameters. Definitely the smart experimentalists will do good jobs to measure them in
the near future to determine whether the bound states exist.

Even though in this work we only deal with the $B\pi$ molecular states, the same approach can be easily applied
to study the $D\pi$ molecular states. Only difference is that the evaluated $D\pi$ mass is smaller than $D^*$, so that
$D\pi$ molecule cannot decay via electromagnetic interaction due to the constraint of the final state phase space, thus
it only has weak decay portals. Definitely, since the rates of weak decays are obviously small, so the measurements on
such $D\pi$ molecular states become even tougher, but not impossible.

If the result of our experimental measurements is positive we would have confidence for existence of
molecular states and know more about their inner structures.

\section*{Acknowledgement}
This work is supported by the National Natural Science Foundation
of China (NNSFC) under the contract No. 11375128 and 11575125.

\appendix
\section{The effective interactions}

\cite{Feng:2011zzb,Meng:2007cx,Cheng:2004ru,Ke:2010aw,Haglin:2000ar}
\begin{eqnarray}
\mathcal{L}_{BB\rho}=&&ig_{BB\rho}[\bar B^0\partial_\mu
B^+\rho^{-\mu}-\partial_\mu\bar B^0 B^+\rho^{-\mu} +\bar
B^0\partial_\mu B^0\rho^{0\mu}-\partial_\mu B^-
B^+\rho^{0\mu}+h.c.]\\ \mathcal{L}_{BB\omega}=&&ig_{BB\omega}[\bar
B^0\partial_\mu B^0\omega^{\mu}-\partial_\mu B^-
B^+\omega^{\mu}+h.c.
]\\\mathcal{L}_{\rho\pi\pi}=&&ig_{\rho\pi\pi}[\partial_\mu \pi^+
\pi^-\rho^{0\mu}+\partial_\mu \pi^0 \pi^+\rho^{-\mu} -\partial_\mu
\pi^+\pi^0\rho^{-\mu}
+h.c.]\\\mathcal{L}_{B^*B\pi}=&&ig_{B^*B\pi}[\partial_\mu\pi^+B^{0*}B^--\pi^+B^{0*}\partial_\mu
B^--\partial_\mu\pi^+B^{0}B^{-*}+\pi^+\partial_\mu B^{0}
B^{-*}+h.c.]\\&&i\frac{g_{B^*B\pi}}{\sqrt{2}}[\pi^0\partial_\mu
B^0\bar B^{0*}-\partial_\mu\pi^0B^0\bar B^{0*}+\partial_\mu\pi^0
B^+B^{-*}-\pi^0\partial_\mu B^+
B^{-*}+h.c.]\\\mathcal{L}_{B^*B\gamma}=&&ig_{B^*B\gamma}e\varepsilon^{\mu\nu\alpha\beta}\partial_\mu
A_\nu(B^*_\alpha\partial_\beta B^\dagger-\partial_\beta B^*_\alpha
B^\dagger+h.c.)
\\\mathcal{L}_{\rho\pi\gamma}=&&ig_{\rho\pi\gamma}e\varepsilon^{\mu\nu\alpha\beta}\partial_\mu
A_\nu(\rho^0_\alpha\partial_\beta \pi^0-\partial_\beta
\rho^0_\alpha \pi^0+\rho^+_\alpha\partial_\beta
\pi^--\partial_\beta \rho^+_\alpha
\pi^-+h.c.)\\\mathcal{L}_{BB\gamma}=&&e A_\mu(\partial_\mu B
B^\dagger- B \partial_\mu B^\dagger)\\\mathcal{L}_{B^*B^*\pi}=&&-
\frac{g_{B^*B^*\pi}}{m_{B^*}}\varepsilon^{\mu\nu\alpha\beta}\partial_{\mu}B^*_{\nu}{
B^*}^\dagger_{\alpha}
\partial_\beta{\pi}\\\mathcal{L}_{B^*B\rho}=&&-
\frac{g_{B^*B\rho}}{m_{B^*}}\varepsilon^{\mu\nu\alpha\beta}
(B\partial_{\mu}\rho_{\nu}\partial_{\alpha}{B_\beta^{*\dagger}}+\partial_{\mu}{B}_\nu\partial_{\alpha}\rho_{\beta}B^\dagger)
\\\mathcal{L}_{BB\sigma}=&&-2m_Bg_\sigma\sigma B
B^\dagger\\\mathcal{L}_{\sigma\pi\pi}=&&-2m_\sigma
g_{\sigma\pi\pi}\sigma\pi \pi^\dagger
\end{eqnarray}


\begin{thebibliography}{99}
\bibitem{Choi:2003ue}
  S.~K.~Choi {\it et al.}  [Belle Collaboration],
  Phys.\ Rev.\ Lett.\  {\bf 91}, 262001 (2003)
  [arXiv:hep-ex/0309032].

\bibitem{Abe:2007jn}
  K.~Abe {\it et al.}  [Belle Collaboration],
  Phys.\ Rev.\ Lett.\  {\bf 98}, 082001 (2007)
  [arXiv:hep-ex/0507019].


\bibitem{Choi:2005} S.~K.~Choi {\it et al.}  [Belle Collaboration],
 Phys.\ Rev.\ Lett.\  {\bf 94}, 182002 (2005).

\bibitem{Choi:2007wga}
  S.~K.~Choi {\it et al.}  [BELLE Collaboration],
  Phys.\ Rev.\ Lett.\  {\bf 100}, 142001 (2008)
  [arXiv:0708.1790 [hep-ex]].





%
\bibitem{Aubert:2005rm}
  B.~Aubert {\it et al.} [BaBar Collaboration],
   Phys.\ Rev.\ Lett.\  {\bf 95}, 142001 (2005)  doi:10.1103/PhysRevLett.95.142001  [hep-ex/0506081].  
\bibitem{Ablikim:2013emm}
  M.~Ablikim {\it et al.} [BESIII Collaboration],
  Phys.\ Rev.\ Lett.\  {\bf 112}, 132001 (2014)
  [arXiv:1308.2760 [hep-ex]].



\bibitem{Ablikim:2013wzq}
  M.~Ablikim {\it et al.}  [BESIII Collaboration],
   Phys.\ Rev.\ Lett.\  {\bf 111}, 242001 (2013)  [arXiv:1309.1896 [hep-ex]].



\bibitem{Ablikim:2013mio}
  M.~Ablikim {\it et al.}  [BESIII Collaboration],
   Phys.\ Rev.\ Lett.\  {\bf 110}, 252001 (2013)  [arXiv:1303.5949 [hep-ex]].



\bibitem{Liu:2013dau}
  Z.~Q.~Liu {\it et al.}  [Belle Collaboration],
  Phys.\ Rev.\ Lett.\  {\bf 110}, 252002 (2013)  [arXiv:1304.0121 [hep-ex]].  
\bibitem{Collaboration:2011gj}
  B.~Collaboration,
  arXiv:1105.4583 [hep-ex].

\bibitem{D0:2016mwd}
  V.~M.~Abazov {\it et al.} [D0 Collaboration],
   Phys.\ Rev.\ Lett.\  {\bf 117}, no. 2, 022003 (2016)  doi:10.1103/PhysRevLett.117.022003  [arXiv:1602.07588 [hep-ex]].  


\bibitem{Xiao:2013iha}
  T.~Xiao, S.~Dobbs, A.~Tomaradze and K.~K.~Seth,
   Phys.\ Lett.\ B {\bf 727}, 366 (2013)  [arXiv:1304.3036 [hep-ex]].  



\bibitem{Deng:2014gqa}
  C.~Deng, J.~Ping and F.~Wang,
   Phys.\ Rev.\ D {\bf 90}, 054009 (2014)  doi:10.1103/PhysRevD.90.054009  [arXiv:1402.0777 [hep-ph]].  

\bibitem{Wang:2014gwa}
  Z.~G.~Wang,
   Eur.\ Phys.\ J.\ C {\bf 74}, 2963 (2014)  doi:10.1140/epjc/s10052-014-2963-7  [arXiv:1403.0810 [hep-ph]].  




\bibitem{Wang:2013cya}
  Q.~Wang, C.~Hanhart and Q.~Zhao,
   Phys.\ Rev.\ Lett.\  {\bf 111}, 132003 (2013)  [arXiv:1303.6355 [hep-ph]].  

\bibitem{Wilbring:2013cha}
  E.~Wilbring, H.-W.~Hammer and U.-G.~Mei$\beta$ner,
  Phys.\ Lett.\ B {\bf 726}, 326 (2013)  doi:10.1016/j.physletb.2013.08.059  [arXiv:1304.2882 [hep-ph]].  
\bibitem{Voloshin:2013dpa}
  M.~B.~Voloshin,
  Phys.\ Rev.\ D {\bf 87}, 091501 (2013)   [arXiv:1304.0380 [hep-ph]].



\bibitem{Cui:2013yva}
  C.~Y.~Cui, Y.~L.~Liu, W.~B.~Chen and M.~Q.~Huang,
  J.\ Phys.\ G {\bf 41}, 075003 (2014)  [arXiv:1304.1850 [hep-ph]].  

\bibitem{Zhang:2013aoa}
  J.~R.~Zhang,
  Phys.\ Rev.\ D {\bf 87}, 116004 (2013)   [arXiv:1304.5748 [hep-ph]].  

\bibitem{Liu:2013vfa}
  X.~H.~Liu and G.~Li,
   Phys.\ Rev.\ D {\bf 88}, 014013 (2013) [arXiv:1306.1384 [hep-ph]].  

\bibitem{Dias:2013fza}
  J.~M.~Dias, F.~S.~Navarra, M.~Nielsen and C.~Zanetti,
  arXiv:1311.7591 [hep-ph].  
\bibitem{Ke:2013gia}
  H.~W.~Ke, Z.~T.~Wei and X.~Q.~Li,
  Eur.\ Phys.\ J.\ C {\bf 73}, 2561 (2013)  [arXiv:1307.2414 [hep-ph]].  

\bibitem{Burns:2016gvy}
  T.~J.~Burns and E.~S.~Swanson,
  Phys.\ Lett.\ B {\bf 760}, 627 (2016)
  doi:10.1016/j.physletb.2016.07.049
  [arXiv:1603.04366 [hep-ph]].

\bibitem{Guo:2016nhb}
  F.~K.~Guo, U.~G.~Mei?ner and B.~S.~Zou,
  Commun.\ Theor.\ Phys.\  {\bf 65}, no. 5, 593 (2016)
  [arXiv:1603.06316 [hep-ph]].

\bibitem{Lu:2016zhe}
  Q.~F.~L¨¹ and Y.~B.~Dong,
  Phys.\ Rev.\ D {\bf 94}, no. 9, 094041 (2016)
  doi:10.1103/PhysRevD.94.094041
  [arXiv:1603.06417 [hep-ph]].

\bibitem{Chen:2016npt}
  X.~Chen and J.~Ping,
  Eur.\ Phys.\ J.\ C {\bf 76}, no. 6, 351 (2016)
  doi:10.1140/epjc/s10052-016-4210-x
  [arXiv:1604.05651 [hep-ph]].




\bibitem{Guo:2007mm}
  X.~H.~Guo and X.~H.~Wu,
  Phys.\ Rev.\  D {\bf 76} (2007) 056004
  [arXiv:0704.3105 [hep-ph]].

\bibitem{Feng:2011zzb}
  G.~Q.~Feng, Z.~X.~Xie and X.~H.~Guo,
  Phys.\ Rev.\  D {\bf 83} (2011) 016003.

\bibitem{PDG12}
  K.~A.~Olive {\it et al.}  [Particle Data Group Collaboration],
  Chin.\ Phys.\ C {\bf 38}, 090001 (2014).



\bibitem{Ke:2012gm}
  H.~W.~Ke, X.~Q.~Li, Y.~L.~Shi, G.~L.~Wang and X.~H.~Yuan,
  JHEP {\bf 1204}, 056 (2012)
  doi:10.1007/JHEP04(2012)056
  [arXiv:1202.2178 [hep-ph]].



\bibitem{Lee:2009hy}
  I.~W.~Lee, A.~Faessler, T.~Gutsche and V.~E.~Lyubovitskij,
  Phys.\ Rev.\ D {\bf 80}, 094005 (2009)
  [arXiv:0910.1009 [hep-ph]].


\bibitem{Meng:2007cx}
  C.~Meng and K.~-T.~Chao,
  Phys.\ Rev.\ D {\bf 75}, 114002 (2007)
  [hep-ph/0703205].





\bibitem{Cheng:2004ru}
  H.~Y.~Cheng, C.~K.~Chua and A.~Soni,
  Phys.\ Rev.\  D {\bf 71}, 014030 (2005)
  [arXiv:hep-ph/0409317].









  H.~W.~Ke, X.~Q.~Li, Z.~T.~Wei and X.~Liu,
  Phys.\ Rev.\  D {\bf 82}, 034023 (2010)
  [arXiv:1006.1091 [hep-ph]].












\bibitem{Ke:2010aw}
  H.~W.~Ke, X.~Q.~Li and X.~Liu,
  Phys.\ Rev.\ D {\bf 82}, 054030 (2010)
  doi:10.1103/PhysRevD.82.054030
  [arXiv:1006.1437 [hep-ph]].
\bibitem{Haglin:2000ar}
  K.~L.~Haglin and C.~Gale,
  Phys.\ Rev.\ C {\bf 63}, 065201 (2001)
  doi:10.1103/PhysRevC.63.065201
  [nucl-th/0010017].






\end{thebibliography}
\end{document}